%% file: Halimi_IEEE_RobustLidar_Sub1.tex
\documentclass[10pt,journal,final,a4paper,twoside,twocolumn]{IEEEtran}

\usepackage{dsfont}
\usepackage{amsmath}
\usepackage{amssymb}
\usepackage{amsfonts}
\usepackage{graphicx}
\usepackage{amsmath}
\usepackage[all,poly]{xy}
\usepackage{multirow}
\usepackage{algorithm}
\usepackage{algorithmic}
\usepackage{color}
\usepackage[noadjust]{cite}
\usepackage{subfigure}
\usepackage{tikz,pgf}
\usetikzlibrary{fit}
\usetikzlibrary{arrows,automata}
\usepackage{verbatim}
\usetikzlibrary{positioning,shapes.geometric}
\usepackage{url}
\usepackage[T1]{fontenc}
\usepackage{hyperref}

\newcommand{\figwidth}{\columnwidth}

\newcommand{\beps}{\boldsymbol{\epsilon}}
\newcommand{\bpsi}{\boldsymbol{\psi}}

\input notations.tex

\input com_notations.tex
\title{Robust and Guided Bayesian Reconstruction of Single-Photon 3D Lidar Data: Application to Multispectral and Underwater Imaging}

\author{Abderrahim Halimi$^{\,1}$\thanks{(1) School of Engineering and Physical Sciences, Heriot-Watt University, Edinburgh, 
EH14 4AS, United Kingdom},  Aurora Maccarone$^{\,1}$, Robert Lamb $^{\,2}$\thanks{(2) Leonardo MW Ltd, Crewe Road North, Edinburgh, EH5 2XS, UK }, Gerald S. Buller $^{\,1}$, Stephen McLaughlin $^{\,1}$\thanks{This work was supported by the UK Royal Academy of Engineering under the Research Fellowship Schemes (RF/201718/17128, RF/201920/19/190), the EPSRC Grants EP/T00097X/1, EP/N003446/1, and by DSTL DASA project DSTLX1000147844. }}

\begin{document}

\maketitle

\begin{abstract}
 
3D Lidar imaging can be a challenging modality when using multiple wavelengths, or when imaging in high noise environments (e.g., imaging through obscurants).
This paper presents a hierarchical Bayesian algorithm for the robust reconstruction of multispectral single-photon Lidar data in such environments.  The algorithm exploits multi-scale information to provide robust depth and reflectivity estimates together with their uncertainties to help with decision making. The proposed weight-based strategy allows the use of available guide information that can be obtained by using state-of-the-art learning based algorithms. The proposed  Bayesian model and its estimation algorithm are validated on both synthetic and real images showing competitive  results regarding the quality of the inferences and the computational complexity when compared to the state-of-the-art algorithms.
   
\end{abstract}


\section{Introduction} \label{sec:Introduction}
Three-dimensional (3D) imaging has generated significant interest from the scientific community due to its increasing use in applications such as  self-driving  autonomous vehicles.
Single-photon light detection and ranging (Lidar) is a technology for high resolution 3D imaging, where its high sensitivity and excellent surface-to-surface resolution can provide rich information on the depth profile and reflectivity of observed targets in challenging imaging scenarios. Single-photon LIDAR operates by emitting picosecond duration laser pulses and collecting the reflected photons using a single-photon sensitive  detector which measures the arrival time of each return photon using a time correlated single-photon counting (TCSPC) system \cite{Buller_STQE2007}.  This results in a collection of X-Y pixels, where a timing histogram of photon counts with respect to their time of flight is constructed for each pixel. 
In the presence of a target with partially reflective or scattering surfaces, the histogram will contain a peak whose amplitude and location are related to the object reflectivity and distance from the sensor. This process can be repeated using different laser wavelengths to obtain a multispectral 3D image of the scene.  

Several practical challenges currently limit the use of Lidar in real world conditions. This paper focuses on some of them and provides a principled  statistical-based solution to improve performance.
Such challenges include the photon sparse regime often observed for long-range imaging \cite{Pawlikowska_OE2017} or rapid imaging based on short acquisition times \cite{LLindell_SigGraph2018,Maccarone_OE19}.  Lidar is also sensitive to the observation environment when imaging in bright conditions \cite{Legros_IEEETIP2021}, and through obscurants or turbid media, such as underwater  \cite{Maccarone_OE19,Halimi_TCI2017b}, or through fog, rain \cite{Wallace_TVT2020}. The latter causes photon scattering which results in the immersion of the useful signal within a high and possibly non-uniform background level \cite{satat2018ICCP,Tobin_OE_2019,Rapp_TCI2017}. To obtain more detailed information about the observed target, one approach is to use multiple laser wavelengths which inevitably lead to larger data volumes which may necessitate the requirement for advanced algorithms to only select  useful pixels \cite{HalimiSSPD2020,TachellaEusipco2020} or to account for shared data structures and correlations \cite{Altmann_TCI2017,Legros_TCI2020,Altmann_SSPD2016,Halimi_TCI2020}.

Several solutions have been proposed in the literature to tackle these challenges.  We distinguish three broad families: statistical, learning-based and hybrid methods. 
The former builds on a statistical model and solves the resulting inference using stochastic simulation methods  \cite{AltmannTIP2015a,Hernandez_MarinPAMI2008,Tachella_TCI2020,Halimi_TCI2017b}, or optimization algorithms \cite{Rapp_TCI2017,Halimi2016Eusipco_a,Halimi_TCI2020}. These principled methods benefit from a good interpretability but are subject to the definition of good features to represent the data.  
The second family learns important features from training data with an available ground-truth, and then uses the learned features to process new measured data \cite{LLindell_SigGraph2018,Ruget_OE2021}. These approaches are dependent on the training data, and might require  expensive network retraining if the imaging conditions change (e.g., different noise level).
The third family uses a plug-and-play (PnP) approach \cite{Venkatakrishnan_CSP_2013} by combining methods of different families to improve performance \cite{Willem_CAMSAP_2017,Tachella_NC_2019}. Beside providing good results, these methods can lack interpretability (e.g., in terms of convergence) and  increasing interest is now devoted to providing principled PnP formulations as in \cite{Yaniv_SAIM_2017,Monga_SPM2020}.
  
This paper proposes a statistical-based algorithm for robust
processing of multispectral 3D Lidar data acquired through obscurants. 
An approximate likelihood distribution is considered and a hierarchical Bayesian model is proposed to exploit the data Poisson statistics, the multi-scale information (known to improve noise and photon-sparsity robustness \cite{Rapp_TCI2017,Tachella_Siam2019,Willem_CAMSAP_2017,Halimi_TCI2020}), and prior knowledge on the depth and reflectivity maps. This hierarchical model ensures the robustness of the proposed strategy to the mismatch between the simplified observation model and the actual one.
The statistical model introduces latent variables that are connected to the parameters of interest using Markov random fields to account for spatial correlations between pixels. This formulation allows for independent parameter updates, which allows for fast parallel implementations. Inspired by the PnP approaches, we propose a weight-based model which exploits the results of state-of-the-art algorithms as a guide to improve performance. The parameter's posterior distribution is obtained by combining the likelihood and proposed prior distributions. This distribution provides parameter estimates together with their uncertainties which are essential for result analysis and decision making. More precisely, we used a coordinate descent algorithm \cite{Bertsekas1995,Sigurdsson2014,HalimiTGRS2015} to approximate the maximum a-posteriori estimator of all parameters, leading to simple iterative updates based on analytical or well known operators (e.g., weighted median filter \cite{Zhang_IEEECVPR_2014,Arce_Book2004}).  The new algorithm is tested on simulated and real underwater data showing promising results in terms of robustness to noise, interpretability and computational cost when compared to state-of-the-art algorithms.
	
The paper is structured as follows. Section \ref{sec:Problem_formulation} introduces the observation model and formulates the considered approximated likelihood. The proposed hierarchical Bayesian model is presented in Section \ref{sec:Bayesian_model}, and the choice of the guidance weights is described in details in Section  \ref{sec:Incorporating_guidance_using_weights_selection}.  Section \ref{sec:estimation_algorithm} then introduces the estimation algorithm used to approximate the maximum a-posteriori estimate of the parameters.  Section \ref{sec:Results_on_simulated_data} analyses the proposed algorithm's performance when considering synthetic data with known ground-truth. Results on real data are presented in Section \ref{sec:Real_3D_underwater_imaging}. Conclusions and future work are finally reported in Section \ref{sec:Conclusions}.

\section{Problem formulation} \label{sec:Problem_formulation}  

This section introduces the observation model for multi-spectral Lidar, followed by the likelihood approximation used in this paper. The last part presents the multi-scale information which is a key ingredient to restore the parameters of interest. 

\subsection{Observation model} \label{subsec:Observation_model}


In addition to object reflectivity, the TSCPC Lidar system measures the depth profiles by illuminating the scene and measuring the time-of-flight of the returned photons. These photons are then collected in a histogram of counts, denoted  $\bsy_{n,t}$, and representing the received photon counts at pixel location $n \in \left\lbrace1,\cdots,N\right\rbrace$, and time-of-flight (ToF) bin $t \in \left\lbrace1,\cdots,T\right\rbrace$. In the case of multi-spectral imaging, the system illuminates the scene using $K$  wavelengths leading to $K$  histograms, denoted by  $\bsy_{n,t,k}$, where $k \in \left\lbrace1,\cdots,K\right\rbrace$.  It is often assumed that the resulting histograms of counts follow a Poisson distribution $\mathcal{P} \left(.\right)$ as follows \cite{Halimi_TCI2020,Rapp_TCI2017,AltmannTIP2015a}: \vspace{-0.2cm} 
\begin{equation}
y_{n,t,k} \sim \mathcal{P} \left(s_{n,t,k}\right)    \vspace{-0.2cm}
\label{eqt:Statistical_model}
\end{equation} 
where $s_{n,t,k} $  represents the average photon counts in the $n$th pixel, $t$th time bin and $k$th wavelength. In presence of at-most one target per-pixel, the signal $ $ can be approximated as follows 
\begin{equation}
s_{n,t,k} = r_{n,k} f_k \left(t-d_{n} \right) + b_{n,t,k}
\label{eqt:MixtureModel}
\end{equation} 
where  $f_k$  represents the system impulse response (SIR) of the $k$th wavelength, which can be measured during system calibration, $r_{n,k}\geq 0$ represents the reflectivity of the observed object assumed different for different wavelengths, $d_{n}\geq 0$ represents the object distance assumed the same for all wavelengths (related to the object depth profile), and  
  $b_{n,t,k} \geq 0$ represents the  background which gathers all photon events that do not originate from reflections at the target surface, i.e., the  dark counts of the detector and the environment background due to the ambient illumination  or photon scattering when imaging through obscurants.  When imaging through turbid media, the background will have a non-uniform shape with respect to the depth observation timing window  \cite{Pfennigbauer_2014,satat2018ICCP},  hence the dependence of $b$  on $t$.  Our goal is to estimate the depth and reflectivity parameters when imaging in extreme conditions due to imaging though obscurants (high and non-uniform background) or sparse photon imaging (e.g., rapid  or long-range imaging).

\subsection{Approximated Poisson likelihood } \label{subsec:Observation_model} 

Assuming independence between the observed pixels $y_{n,t,k}$  leads to the joint likelihood \vspace{-0.2cm}
\begin{equation}
P(\bsY | \bsd,\bsR,\bsB) = \prod_{n=1}^{N} \prod_{k=1}^{K} \prod_{t=1}^{T}{ \frac{s_{n,t,k}^{y_{n,t,k}}}{y_{n,t,k}!}  \exp^{-s_{n,t,k}}  } 
\label{eqt:Likelihood}
\end{equation}
where  $\bsd$  is an $N \times 1$    vector gathering depth values, $\bsR, \bsB$  are $N \times K$    matrices gathering reflectivity and background values, respectively. 
In absence of background counts and  assuming that  $ \sum_{t=1}^{T}{f_k\left(t-d_{n} \right)} = 1, \forall k$  for all realistic $d_{n}$, the likelihood reduces to 
\begin{eqnarray}
P({}^s\!\bsy_{n} | \bsr_{n}, d_{n})   & \propto &
\prod_{k=1}^{K} \left[ \mathcal{G} \left(r_{n,k}; 1+\bar{s}_{n,k},  1 \right)  \bar{Q}\left({}^s\!\bsy_{n,k}\right) \right]
\nonumber \\
& \times & \prod_{t,k} f_k \left[ \left(t-d_{n} \right)\right]^{{}^s\!y_{n,t,k}} 
\label{eqt:Likelihood3} 
\end{eqnarray} 
where $  \propto$ stands for proportional to, $\mathcal{G} (x; .,.)$ denotes the gamma distribution with shape and scale parameters,  ${}^s\!\bsy_{n,k}$ represents the histogram of target reflected (or signal) counts,   $\bar{Q}\left({}^s\!\bsy_{n,k}\right)$ is a function of signal counts and 
\begin{equation}
r_{n,k}^{\textrm{ML}} = \bar{s}_{n,k} =\sum_{t=1}^{T}   {}^s\! y_{n,t,k}
\label{eqt:r_ML}
\end{equation} 
represents the maximum likelihood (ML) estimate of the reflectivity at the $k$th wavelength obtained by summing the signal counts. Note that the depth maximum likelihood estimate is obtained using a simple log-matched filtering of the histogram with the SIR, as follows
\begin{equation}
{d}^{\textrm{ML}}_{n} = \textrm{argmax}_{d} \sum_{t, k}{  {}^s\! y_{n,t,k} \textrm{log}[f_k(t - d)] }.  \label{eqt:D_ML}
\end{equation} 
It is common to approximate the SIR at each wavelength with the Gaussian function    $f_k \left(\mu - d_{n} \right)   =  \mathcal{N}(d_{n};  \mu. \sigma_k^2) $  \cite{Halimi2016Eusipco_a,Altmann_TIP2020}.  In this case the likelihood in \eqref{eqt:Likelihood3}  becomes
\begin{eqnarray}
P({}^s\!\bsy_{n} | \bsr_{n}, d_{n})   & \propto &
\prod_{k=1}^{K} \left[ \mathcal{G} \left(r_{n,k}; 1+\bar{s}_{n,k},  1 \right)  \bar{Q}\left({}^s\!\bsy_{n,k}\right) \right]
\nonumber \\
& \times & 
{\mathcal{N}(d_{n}; {d}^{\textrm{ML}}_{n} , \bar{\sigma}^2      )  } 
\label{eqt:Likelihood4} 
\end{eqnarray} 
where $ \mathcal{N}(x;  \mu. \sigma_k^2) $ represents the Gaussian distribution with average $\mu$ and variance $\sigma_k^2$,  $\bar{\sigma}^{2}  =  \left( {\sum_{k}  \frac{\bar{s}_{n} } {\sigma_k^2} }\right)^{-1}$ and   
${d}^{\textrm{ML}}_{n}=   \bar{\sigma}^2  \sum_{k=1}^{K} \frac{\sum_{t=1}^{T}  t  {}^s\! y_{n,t,k}}{\sigma_k^2}$  is given analytically when considering Gaussian approximation for the SIR.  
Considering these approximations, Eq. \eqref{eqt:Likelihood4}  indicates that the depth and reflectivity parameters are independent and that they appear within conventional Gaussian and gamma distributions, which is crucial for the design of the proposed Bayesian strategy. Indeed, the quality of the ML depth and reflectivity estimators is known to be poor in challenging scenarios, hence the need to account for known parameter properties to improve reconstruction. This can be done within the Bayesian framework adopted in this paper.

\subsection{Multiscale information} \label{subsubsec:Multiscale_information} 

A common approach to improve the performance of maximum likelihood estimation for Lidar data is to consider multi-scale information, as already exploited in several state-of-the-art 3D Lidar denoising algorithms \cite{Rapp_TCI2017,Tachella_Siam2019,Willem_CAMSAP_2017,Halimi_TCI2020}.
The key observation is that spatially downsampled histograms, which are still Poisson distributed, lead to depth and reflectivity estimates with lower noise at a price of a reduced spatial resolution. 
In this paper, we adopt a similar strategy by considering $L$ downsampled version of the histogram of counts. 
For each wavelength $k$, spatially downsampled version of the histograms $\bsY$ are first computed based on predefined $L$ graphs of neighbours  $\phi^{1,\cdots,L}$ leading to  $\bsY^{\ell}_{k}$ (for example,  $q^{(2)} = 3\times 3$ neighbours for $\phi^{(2)}$, a $q^{(3)}=5\times 5$ neighbours for $\phi^{(3)}$, $\cdots$). The latter can be efficiently computed using convolutions in the case of a regular grid but our algorithm can be equally applied to a non-uniform sampling grid of the pixels.  Assuming independence between these histograms lead to $L$ likelihood distributions as follows
\begin{eqnarray}
P({}^s\!\bsy_{n}^{(\ell)} | \bsr_{n}^{(\ell)}, d_{n}^{(\ell)})  & \propto & \prod_{k=1}^{K}  \left[
\mathcal{G} \left(r_{n,k}^{(\ell)}; 1+\bar{s}_{n,k}^{(\ell)},  1 \right)
\bar{Q}\left({}^s\!\bsy_{n,k}^{(\ell)}\right) \right]
\nonumber \\
& \times & {\mathcal{N}( d_{n}^{(\ell)}; {d}_{n}^{\textrm{ML}(\ell)}, \bar{\sigma}^{2 (\ell)} )  }
\label{eqt:Likelihood4b} 
\end{eqnarray}
$\forall \ell \in {1,\cdots,L}$, where  $\bar{\sigma}^{2 (\ell)}  =  \left( {\sum_{k}  \frac{\bar{s}_{n}^{(\ell)}} {\sigma_k^2} }\right)^{-1}$,   $\ell=1$ is the original cube, and for example,  $\ell=2$ corresponds to a  $3\times 3$ downsampling,  $\ell=3$ to a $5\times 5$  downsampling, etc.

\section{Hierarchical Bayesian model} \label{sec:Bayesian_model} 

Estimating depth and reflectivity parameters in extreme conditions is an  ill-posed problem which requires the use of prior information to alleviate its indeterminacy. A Bayesian strategy is considered to combine the approximate likelihood described above, with parameter prior distributions accounting for known parameter properties.  The resulting posterior distribution will be exploited by deriving Bayesian point estimators and additional measures of uncertainty about the estimates. The following sub-sections introduce the  proposed Bayesian model.

\subsection{Prior distribution for depth} \label{subsec:D_Prior_distribution}
  
Our model assumes the observation of $L$  depth maps $\bsd^{(\ell)}$ obtained from multi-scale downsampled histograms, and having different noise levels as highlighted by the Gaussian variances in \eqref{eqt:Likelihood4b}. 
Object depth profiles exhibit homogeneous surfaces (i.e., spatial correlation) separated by a discontinuous jump between different surfaces. This requires enforcing spatial correlation between the pixels of a surface, while preserving edges of isolated objects or between separated surfaces. 
To mitigate this information, we introduce an $N \times 1$  latent variable $\bsx$ that is connected  to all multi-scale depth maps, to provide a robust reconstruction of the true depth map by considering correlations between pixels. To preserve edges separating different surfaces, we propose the following mixture of Laplace conditional prior distributions for $\bsx$ as follows 
\begin{eqnarray}
x_{n} | d_{n}^{(1,\cdots, L)}, w_{\nu_n,n}^{(1,\cdots, L)},  \epsilon_{n} \sim  \nonumber \\
 \prod_{n' \in \nu_{n}}  \left[ \prod_{\ell=1 }^{L}  {\mathcal{L}\left(x_{n};  d_{n'}^{(\ell)},   \frac{\epsilon_{n}}{ w_{n',n}^{(\ell)}   }  \right)  } \right]  
  \label{eqt:Gaussian_Prior} 
\end{eqnarray}
where $ \mathcal{L}(x;  \mu. \epsilon) $ represents the Laplace distribution with average $\mu$ and variance or diversity parameter $\epsilon$, 
 $ \nu_{n}$ represents the spatial neighbourhood of the $n$th pixel, $ d_{n'}^{(\ell)}$ denotes the mean,   $\epsilon_{n}>0$ is the variance of $\bsx_{n}$ and $w_{n',n}^{(\ell)} \geq 0$ are constant weights to be defined. Note that   \eqref{eqt:Gaussian_Prior} 
 preserves edges as it considers the sparsity promoting $\ell_1$-norm of the differences between $\bsx$ and $\bsd$.  The weights  $w_{n',n}^{(\ell)} \geq 0$ are essential as they allow guiding the connections between $\bsx$ and $\bsD$ using any available side-information (e.g., obtained from other sensors in the case of multi-modal imaging, or by using state-of-the-art denoising algorithms in the case of plug-and-play approaches).  
  It is also worth noting that prior \eqref{eqt:Gaussian_Prior}  is connected to the Bayesian lasso model \cite{Park_JASA_2008,Figueiredo_TPAMI_2003}. Indeed,  \eqref{eqt:Gaussian_Prior}  could be obtained by marginalizing the exponentially-distributed variance hyper-parameter  of a Gaussian mixture prior. 
Finally, \eqref{eqt:Gaussian_Prior}  does not enforce positivity on the depth parameter $\bsx$, however, this will be ensured as indicated in Section \ref{subsec:Depth_estimation}.

\subsection{Prior distribution for reflectivity} \label{subsec:R_Prior_distribution} 
In a similar fashion to depth, spatial smoothness can be enforced on the reflectivity by considering latent variables as in the gamma Markov random field prior \cite{DikmenTASLP2010}. However, this prior will lead to underestimated reflectivity values as already highlighted in \cite{Tachella_TCI2020}.  In this work,  we introduce an $N \times K$ latent variable $\bsM$  assigned a Gaussian prior distribution as follows  
\begin{eqnarray}
 m_{n,k}  |   r_{\nu_n,k}^{(1,\cdots, L)}, v_{\nu_n,n}^{(1,\cdots, L)},  \psi_{n,k}^2 \sim  \nonumber \\ 
 \prod_{n' \in \nu_{n}}  \left[ \prod_{\ell=1 }^{L}  {
\mathcal{N} \left(r_{n',k}^{(\ell)}, 
 \frac{\psi^2_{n,k}}{ v_{n',n; k}^{(\ell) }  } \right) 
} \right] 
  \label{eqt:Gaussian_Prior_M} 
\end{eqnarray}
where  $v_{n',n;k}^{(\ell)} \geq 0$ are constant weights to be defined, and $\psi^2_{n,k}$ represents the variance of the latent variable and contains reflectivity uncertainty information for the $k$th wavelength. The variable $\bsm_k$ contains multi-scale reflectivity information through  $ r_{n,k}^{(\ell)}$ and  will serve as the reflectivity estimate for the  $k$th wavelength. 
 Although this is not a conjugate prior, it will lead to non-negative analytical estimates for $\bsM, \bsR$ as indicated in Section \ref{sec:estimation_algorithm}.

\subsection{Priors of the variance hyperparameters } \label{subsec:Hyperparameters} 
 
The  variance parameters $\epsilon_{n}, \forall n$ (resp. $\psi_{n,k}^2,  \forall n,k$) should be positive. Assuming  prior independence between the parameters $\epsilon_{n}, \forall n$ (resp. $\psi_{n,k}^2, \forall n,k$) and accounting for their  positivity, we assign a conjugate inverse gamma distribution for these parameters as follows
\begin{eqnarray}
f\left(\beps\right)  = \prod_{n=1}^{N}  \mathcal{IG} \left({ \epsilon_{n}}; \alpha_d, \beta_d \right) \\
f\left(\bpsi \right)  = \prod_{k=1}^{K}  \prod_{n=1}^{N}  \mathcal{IG} \left( \psi_{n,k}^2; \alpha_r, \beta_r \right) 
\label{eqt:Variance} 
\end{eqnarray} 
where $\alpha_r, \beta_r, \alpha_d, \beta_d$ are positive user fixed hyperparameters. In absence of additional knowledge, these hyperparameters are fixed to obtain a non-informative prior.


\subsection{Posterior distribution} \label{subsec:Posterior_distribution} 

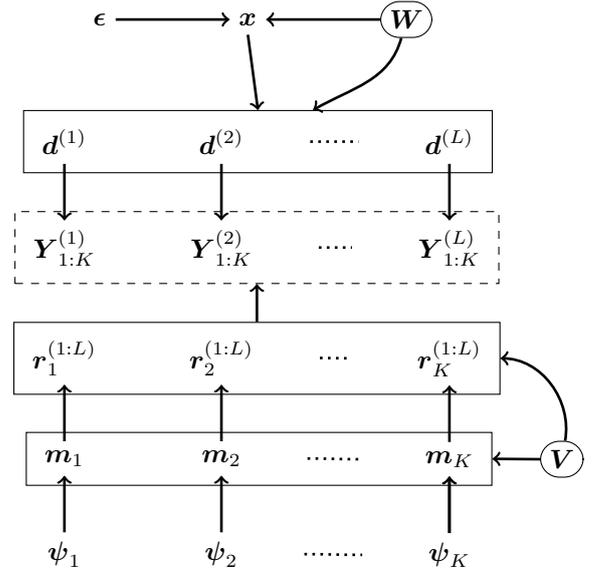
\begin{figure}[h!]
\centering
\begin{tikzpicture}
 nodes %
\node[text centered] (Y1) {$\bsY^{(1)}_{1:K}$};
\node[right =1.  of Y1, text centered] (Y2) {$\bsY^{(2)}_{1:K}$};
\node[right =0.5  of Y2, text centered] (Y25) {};
\node[right =4.  of Y1, text centered] (YL) {$\bsY^{(L)}_{1:K}$};
\node[left =0.5  of YL, text centered] (YL5) {};

\node[above =0.75  of Y1, text centered] (d1) {$\bsd^{(1)}$};
\node[above =0.75  of Y2, text centered] (d2) {$\bsd^{(2)}$};
\node[right =0.5  of d2, text centered] (d25) {};
\node[above =0.75  of YL, text centered] (dL) {$\bsd^{(L)}$};
\node[left =0.5  of dL, text centered] (dL5) {};

\node[below =0.75  of Y1, text centered] (r1) {$\bsr^{(1:L)}_{1}$};
\node[below =0.75  of Y2, text centered] (r2) {$\bsr^{(1:L)}_2$};
\node[right =0.5  of r2, text centered] (r25) {};
\node[below =0.75 of YL, text centered] (rL) {$\bsr^{(1:L)}_K$};
\node[left =0.5  of rL, text centered] (rL5) {};

\node[below =0.75  of r1, text centered] (m1) {$\bsm_{1}$};
\node[below =0.75  of r2, text centered] (m2) {$\bsm_2$};
\node[right =0.5  of m2, text centered] (m25) {};
\node[below =0.75 of rL, text centered] (mL) {$\bsm_K$};
\node[left =0.5  of mL, text centered] (mL5) {};
\node[draw, rectangle, right =0.75 of mL, text centered,rounded corners=.25cm] (V) {$\bsV$};

\node[below =0.75  of m1, text centered] (Psi1) {$\bpsi_{1}$};
\node[below =0.75  of m2, text centered] (Psi2) {$\bpsi_2$};
\node[right =0.5  of Psi2, text centered] (Psi25) {};
\node[below =0.75 of mL, text centered] (PsiL) {$\bpsi_K$};
\node[left =0.5  of PsiL, text centered] (PsiL5) {};


\node[above =1.2  of d1, text centered] (UpX) {};

\node[right  =2.05 of UpX, text centered] (Xk) {$\bsx$};

\node[draw, rectangle, right =1.5  of Xk, text centered,rounded corners=.25cm] (W) {$\bsW$};
\node[ left =1.5 of Xk, text centered] (Eps) {$\beps$};
\node[above =0.75  of UpX, text centered] (UpX2) {};



 edges %
\draw[->, line width= 1] (d1) to  [out=270,in=90, looseness=1] (Y1);
\draw[->, line width= 1] (d2) to  [out=270,in=90, looseness=1] (Y2);
\draw[->, line width= 1] (dL) to  [out=270,in=90, looseness=1] (YL);


\draw[->, line width= 1] (m1) to  [out=90,in=270, looseness=1] (r1);
\draw[->, line width= 1] (m2) to  [out=90,in=270, looseness=1] (r2);
\draw[->, line width= 1] (mL) to  [out=90,in=270, looseness=1] (rL);

\draw[->, line width= 1] (Psi1) to  [out=90,in=270, looseness=1] (m1);
\draw[->, line width= 1] (Psi2) to  [out=90,in=270, looseness=1] (m2);
\draw[->, line width= 1] (PsiL) to  [out=90,in=270, looseness=1] (mL);
\draw[->, line width= 1] (PsiL) to  [out=90,in=270, looseness=1] (mL);



\draw[dotted, line width= 1] (Y25) to  [out=0,in=180, looseness=1] (YL5);
\draw[dotted, line width= 1] (d25) to  [out=0,in=180, looseness=1] (dL5);
\draw[dotted, line width= 1] (r25) to  [out=0,in=180, looseness=1] (rL5);
\draw[dotted, line width= 1] (m25) to  [out=0,in=180, looseness=1] (mL5);
\draw[dotted, line width= 1] (Psi25) to  [out=0,in=180, looseness=1] (PsiL5);

\node[draw,dashed,fit=(Y1) (Y2) (YL)] (BlcY){};
\node[draw,rectangle,fit=(d1) (d2) (dL)](BlcD) {};
\node[draw,rectangle,fit=(r1) (r2) (rL)](BlcR) {};
\node[draw,rectangle,fit=(m1) (m2) (mL)](BlcM) {};

\draw[->, line width= 1] (BlcR) to  [out=90,in=270, looseness=1] (BlcY);
\draw[->, line width= 1] (Xk) to  [out=270,in=90, looseness=0] (BlcD);
\draw[->, line width= 1] (W) to  [out=255,in=30, looseness=1] (BlcD);

\draw[->, line width= 1] (V) to  [out=80,in=360, looseness=1] (BlcR);
\draw[->, line width= 1] (V) to  [out=180,in=360, looseness=1] (BlcM);

\draw[->, line width= 1] (W) to  [out=180,in=0, looseness=1] (Xk);
\draw[->, line width= 1] (Eps) to  [out=0,in=180, looseness=1] (Xk);

\vspace{-0.5cm}
\end{tikzpicture}
\caption{DAG for the observations, parameter and hyperparameters of the proposed model. The guiding weights appear in circles,  the observations in a dashed box and the rectangular box gathers the joint multi-scale or multi-wavelengths parameters.}
\label{fig:DAG}
\end{figure}

 The joint posterior distribution of this Bayesian model can be computed from the following hierarchical structure (after dropping indices for clarity)
\begin{eqnarray}
f\left(\bsx, \bsD,  \bsM, \bsR,  \beps,  \bpsi  | \bsY \right)  \propto 
  f(\bsY | \bsR, \bsD)  f(\bsD, \bsx | \beps, \bsW) \nonumber \\
   f(\bsR, \bsM | \bpsi, \bsV)     f(\beps)  f(\bpsi).
\label{eqt:Joint_Posterior}
\end{eqnarray}
This distribution contains complete information regarding the parameters of interest  $\bsx, \bsD, \bsR, \bsM$  and their uncertainties $\beps ,  \bpsi $. 
A common approach is to extract Bayesian point estimators such as the  maximum a-posteriori (MAP) estimator or  the  minimum mean square estimator (MMSE).  In this paper, we consider the MAP estimator of all parameters. It should be noted that the depth related parameters $ \bsD, \bsx, \beps$ and the reflectivity ones  $\bsR, \bsM, \bpsi$ are independent allowing parallel optimization with respect to both set of parameters. Finally,  Fig. \ref{fig:DAG}  presents a directed acyclic graph (DAG) which summarizes the main parameters of the proposed hierarchical Bayesian model.

\section{Incorporating guidance using weights selection} \label{sec:Incorporating_guidance_using_weights_selection} 
 
The choice of the weights is very important and will have a direct impact on the algorithm performance. Several strategies have been considered in the literature where the choice can be based on the spatial distance between points, similarity of their values, etc \cite{Liu_TIP_2017,Chan_SR_2019,Liu_TIP_2019}. In this paper, we assume the presence of guiding information (e.g., by using other algorithms, or sensors) and define these weights while considering multi-scale and multi-wavelength information.  

\subsection{Depth weights $\bsW$} \label{subsec:Depth_Weights} 

Assuming the presence of an outlier free multi-scale guiding depth $\underline{\bsd}^{(\ell)}, \ell=1, \cdots, L$, our selection of the multi-scale weights $\bsW$ encourages the depth map at a given scale ${\bsd}^{(\ell)}$ to be close to $\underline{\bsd}^{(\ell)}$ (with a graph of neighbours  $\phi^{0}$, for example $3 \times 3$ neighbours in a uniform grid).  
More precisely, we assign low weights for pixels  that differ significantly from their corresponding pixels in $\underline{\bsd}^{(\ell)}$ as follows 
\begin{eqnarray} 
w_{n,n'}^{(\ell)}  & = & w_{\textrm{norm}}   \left[ \prod_{\ell'=1}^{\ell-1} \left(1-  w_{n,n'}^{(\ell')} \right)  \right]   \nonumber \\
& \times & \exp\left(   -\frac{ |{d}_n^{\textrm{ML} (\ell)}  - \underline{d}_{n'}^{(\ell)}  |      }{2 \zeta  q^{(\ell)}}    
\right)
\label{eqt:Scale_Weights_W}
\end{eqnarray}
for  $\ell \in \left\lbrace 1,\cdots,L \right\rbrace$,  
where $w_{\textrm{norm}}  $ is a normalization constant ensuring $\sum_{\ell,n'} w_{n,n'}^{(\ell)} =1$,    the coefficient $\zeta$ is easily fixed based on physical considerations related to the impulse response width and it is weighted by  $q^{(\ell)}$ to account for the multi-resolution effect. In \eqref{eqt:Scale_Weights_W}, the product over $\ell'$ promotes lower scales data if their weights are high.

We are now left with the task of finding a reliable multi-scale depth guide which is robust to outliers. This information can be obtained by considering other sensing modalities such as Radar, Sonar, when available. It can also be obtained by applying an off-the-shelf depth reconstruction algorithm to the Lidar data (e.g., \cite{Rapp_TCI2017}).  The latter strategy is adopted in this paper. We consider two methods, the first, denoted GD1  for guide depth 1, is inspired  by  \cite{Shin_TCI2015} which adopted the rank order mean approach to unmix signal from background counts. Here, we first detects background corrupted pixels (those without $\sqrt{q^{(2)}}$ neighbours having close depth values) and then replace them with the median of surrounding valid points. The second strategy, denoted GD2, represents ${\bsd}^{\textrm{ML} (\ell=1:L)}$ as a point cloud and applies an outlier rejection algorithm to remove corrupted values (i.e., using \textit{pcdnoise} in Matlab \cite{Radu2008}).  
We note finally that the weights could be updated with iterations leading to a pseudo-Bayesian approach \cite{Milanfar_TSPM_2013}, but this is out of the scope of this paper and will be left for future work.

\subsection{Reflectivity weights $\bsV$} \label{subsec:Depth_Weights} 

The reflectivity weights are obtained from the multi-scale images $\bsr_k^{(\ell)}, \forall k, \ell$, but we note that they can also be learned using additional reflectivity maps acquired by other sensors when available. Assuming the presence of  $\underline{\bsr}_k^{(\ell)}, \forall k, \ell$ reflectivity guides and depth weights $\bsW$, we consider a multi-scale bilateral filtering approach \cite{Tomasi_ICCV_1198,Chan_SR_2019,Tachella_NC_2019}  and define the reflectivity weights as follows
\begin{equation} 
v_{n,n';k}^{(\ell)} = v_{\textrm{norm}} \,
w_{n,n'}^{(\ell)} \,
 \exp\left(   -\frac{ |r_{n,k}^{\textrm{ML} (\ell)}  - \underline{r}_{n',k}^{(\ell)} |}{2 \eta_{n,k} q^{(\ell)}}    
\right)
\label{eqt:Scale_Weights_V}
\end{equation}
where $v_{\textrm{norm}}$ is a normalization constant ensuring $\sum_{\ell,n'} v_{n,n';k}^{(\ell)} =1$, and $\eta_{n,k}$ is a constant.  As indicated in \eqref{eqt:Scale_Weights_V}, correlation between depth and reflectivity images is introduced through the use  of $\bsW$ to define $\bsV$. 
The reflectivity variables $\bsr_k^{(\ell)}, \forall k, \ell$, follow a gamma distribution and hence show data dependent noise levels. To account for this effect, we assume a signal dependent variance $\eta_{n,k}$, which is fixed as follows 
\begin{equation} 
\eta_{n,k} = \textrm{max} \left(0.1, {r}_{n,k}^{\textrm{ML} (L)}\right).
\label{eqt:Eta}
\end{equation}
 
Several reflectivity restoration algorithms can be used to obtain the guides $\underline{\bsr}_k^{(\ell)}, \forall k, \ell$, based on the considered imaging scenarios. Algorithms based on Poisson statistics can be used in the sparse photon regime \cite{Azzari_SPL2016,Salmon_JMIV2014,Halimi2016Eusipco_a},  while other state-of-the-art denoising algorithms \cite{DabovTIP2007,Zhang_TIP2017} can be considered in dense photon regimes. In this paper, we consider three guidance methods. The first guidance intensity  (denoted GI1) considers  $\underline{\bsr}_k^{(\ell)} = \bsr_k^{\textrm{ML} (\ell)}, \forall k, \ell$
which leads to a multi-scale generalization of the bilateral filter. Indeed, these multi-scale maps already contain filtering properties which will provide good performance in practice.   The second guidance (GI2) considers the Poisson based  reconstruction method  \cite{Azzari_SPL2016} (used with authors defaults parameters) which is applied to each scale and wavelength of $\bsr_k^{\textrm{ML} (\ell)}, \forall k, \ell$ to obtain $  \underline{\bsr}_k^{(\ell)}, \forall k, \ell$. As a third guidance (GI3), we considered the state-of-the-art learning based DnCNN denoiser \cite{Zhang_TIP2017}, also applied to each scale and wavelength $\bsr_k^{\textrm{ML} (\ell)}, \forall k, \ell$.  
We finally note that  reflectivity multi-spectral correlations are introduced through the depth weights, which are shared between all wavelengths. Additional correlations can be easily included through the weights $\bsV$ when building the reflectivity guides.

\section{Estimation algorithm} \label{sec:estimation_algorithm} 

We propose to use the MAP estimators for all parameters and hyperparameters   $\bsx, \bsD, \bsR, \bsM,  \beps, \bpsi$. More precisely, the maximum of the posterior distribution in  \eqref{eqt:Joint_Posterior}   is approximated using a coordinate descent algorithm  \cite{Bertsekas1995,Sigurdsson2014}. This algorithm sequentially maximizes the conditional distributions associated with each parameter until convergence to a local minimum of the negative log-posterior.   
The algorithm's main steps are presented  in Algo. \ref{alg:Denoising_part} 
and described with more details in the following sections. 
Note that the resulting depth updates alternates between robust to outliers non-linear parameter estimation (line 11) and a filtering step (line 12), which are commonly observed steps in several state-of-the-art algorithms  \cite{Yaniv_SAIM_2017,Venkatakrishnan_CSP_2013,Tachella_NC_2019} and optimization algorithms \cite{Boyd_FTML2011}.
Note also that reflectivity and depth iterates are independent and can be run in parallel. Note finally that reflectivity updates are analytically obtained ensuring fast estimation.
\begin{algorithm}
\caption{Estimation algorithm} \label{alg:Denoising_part}
\begin{algorithmic}[1]
       \STATE \underline{Input:}
       \STATE $\bsY_{k}, \forall k; L; \phi^{1,\cdots,L}$
       \STATE \underline{Generate low resolution data:}
			 \STATE Generate low-resolution histograms $\bsY^{(\ell)}_{k}, \ell \in \left\lbrace 1,\cdots,L \right\rbrace $ using $\phi^{1,\cdots,L}$
			 \STATE Estimate background level $\hat{\bsB}_k, \forall k$ as in \eqref{eqt:b_tot} 
			 \STATE Estimate ${\bsd}^{\textrm{ML} (\ell)}, {\bsr}_k^{\textrm{ML} (\ell)}, \forall k$ as in \eqref{eqt:D_ML}, \eqref{eqt:r_ML}  
			 \STATE Compute guiding weights $\bsW, \bsV$ as in \eqref{eqt:Scale_Weights_W}, \eqref{eqt:Scale_Weights_V}
		\STATE \underline{Coordinate descent algorithm}	  
       \WHILE{conv$=0$}
               \STATE Update $x_{n}, \forall n$ using WMF in \eqref{eqt:WeightedMedian_Depth}
               \STATE Update $\bsd^{(\ell)}, \forall \ell$ using threshold operator in \eqref{eqt:Gene_SoftThresh_Depth} 
               \STATE Update $\beps$ using analytical mode in  \eqref{eqt:Mode_posterior_sigma_nk}
                              \STATE Update $\bsm_{k}, \forall k$ using analytical mode in \eqref{eqt:updateM} 
               \STATE Update $\bsr_k^{(\ell)}, \forall \ell$ using analytical mode in  \eqref{eqt:R_update} 
               \STATE Update $\bpsi$ using analytical mode in  \eqref{eqt:Mode_posterior_psi_nk}
               \STATE Set conv$=1$ if the convergence criteria are satisfied
       \ENDWHILE
   
		\STATE \underline{Output:}
			 \STATE $\bsx, \bsM,  \beps, \bpsi$  
\end{algorithmic}
\end{algorithm}

\subsection{Updating $\bsx$ } \label{subsec:Depth_estimation} 
The parameters  of $\bsx$ are independent allowing parallel updating of $x_{n}, \forall n$.  
Minimizing the negative-log of the conditional distribution reduces to  
\begin{equation} 
\hat{x}_{n} = \operatornamewithlimits{\textrm{argmin}}\limits_{x}  
\mathcal{C} (x)  = \operatornamewithlimits{\textrm{argmin}}\limits_{x}   \sum_{\ell, n'\in \nu_{n}}   {w_{n',n}^{(\ell)}} | x - d_{n'}^{(\ell)}|.
\label{eqt:WeightedMedian_Depth}
\end{equation}
This is a weighted median filter (WMF) which has several efficient implementations (e,g,  \cite{Zhang_IEEECVPR_2014}).  
Note that the solution of \eqref{eqt:WeightedMedian_Depth} will be non-negative  provided that $d_{n'}^{(\ell)} \geq 0$, which is ensured during initialization.

\subsection{Updating $\bsD$} \label{subsec:Depth_estimation} 
The variables  $d_1^{(\ell)}, \cdots, d_N^{(\ell)}$  are  independent and spatial correlation is introduced through the latent variable $\bsx$. This is interesting as it allows the parallel implementation of $d_n^{(\ell)}, \forall n, \ell$ with respect to $n$ and $\ell$. Straightforward computations show that the update  $d_n^{(\ell)}$ is given by
\begin{equation} 
d_n^{(\ell)} = \operatornamewithlimits{\textrm{argmin}}\limits_{d}   \frac{   \left[ d - {d}_{n}^{\textrm{ML}(\ell)} \right]^2}{2 \bar{\sigma}^{2 (\ell)} }  + \sum_{n'\in \nu_{n}}  \frac{ w_{n,n'}^{(\ell)}  | d - x_{n'}| } {\epsilon_{n'}^2}.
\label{eqt:Gene_SoftThresh_Depth} 
\end{equation}
This is a generalization of the well known soft-threshold operator which can be  efficiently solved. Note that the solution of \eqref{eqt:Gene_SoftThresh_Depth} will be non-negative  provided that $x_{n'} \geq 0$ and ${d}_{n}^{\textrm{ML}(\ell)} \geq 0$ which is ensured during initialization. 

\subsection{Updating depth variance: $\beps$} \label{subsec:Depth_variance} 
The conditional distribution of $\epsilon_n$ is an inverse-gamma distribution given by 
\begin{equation}
\epsilon_n| \bsx,\bsD, \bsW  \sim   \calI \calG \left[L+\bar{N}  + \alpha_d,  \mathcal{C} \left(x_{n} \right) + \beta_d \right]
\label{eqt:posterior_sigma_nk}
\end{equation}
whose mode is given by  
\begin{equation}
\hat{\epsilon}_{n} =  \frac{\mathcal{C} \left(x_{n} \right) + \beta_d}{L+\bar{N}+ \alpha_d+1}
\label{eqt:Mode_posterior_sigma_nk}
\end{equation}
where $\bar{N}$ is the number of spatial neighbours.

\subsection{Updating $\bsM$ } \label{subsec:M} 
The conditional distribution of $\bsM$ is a normal distribution whose mean is analytically given by
\begin{equation}
\hat{m}_{n,k} = \frac{\sum_{\ell,  n'\in \nu_{n}} v_{n',n; k}^{(\ell) } \, r_{n',k}^{(\ell)}}{\sum_{\ell, n'\in \nu_{n}} v_{n',n; k}^{(\ell) }}. 
\label{eqt:updateM}
\end{equation}
This equation highlights a weighted sum of the multi-scale reflectivity maps $\bsr$, as for the bilateral filter.

\subsection{Updating $\bsR$ } \label{subsec:R} 
The parameters  of $\bsR$ are independent allowing parallel updating of $r_{n,k}^{(\ell)}, \forall n,k,\ell$. 
Minimizing the negative-log of the conditional distribution reduces to  
\begin{equation} 
\hat{r}_{n,k}^{(\ell)} = \operatornamewithlimits{\textrm{argmin}}\limits_{r} \left\lbrace r - \bar{s}_{n,k}^{(\ell)} \textrm{log} r + \mathcal{H}(r)  \right\rbrace
\label{eqt:condPost_R}
\end{equation}
where $\mathcal{H}(r)  = \frac{1}{2 \psi_r }
 \left(r - \mu_r \right)^2$  with 
$\psi_r^{(-1)} = \sum_{n'} \frac{v_{n',n; k}^{(\ell)} } { \psi_{n',k}}$ and $\mu_r  = \sum_{n'}  \left(\frac{v_{n,n'; k}^{(\ell)} m_{n',k}} {\psi_{n',k}}  \right)$.
The minimum is analytically provided by \cite{FigueiredoBioucas-Dias2010}
\begin{equation} 
\hat{r}_{n,k}^{(\ell)} =  \frac{\mu_r - \psi_r + \sqrt{\left(\mu_r - \psi_r\right)^2 + 4 \psi_r \bar{s}_{n,k}^{(\ell)} }} {2}.
\label{eqt:R_update}
\end{equation}

\subsection{Updating $\bpsi$ } \label{subsec:Psi} 
The conditional distribution of the  reflectivity variance $\psi_{n,k}$ is an inverse-gamma distribution given by 
\begin{equation}
\psi_{n,k} | \bsM,\bsR, \bsV  \sim   \calI \calG \left[\frac{L+\bar{N}}{2}  + \alpha_r,  \mathcal{K} + \beta_r \right]
\label{eqt:posterior_psi_nk}
\end{equation}
with $\mathcal{K}= \sum_{\ell, n'\in \nu_{n}}  \frac{ {v_{n',n,k}^{(\ell)}} \left( m_{n,k} - r_{n',k}^{(\ell)} \right)^2}{2}$. The mode is analytically given by  
\begin{equation}
\hat{\psi}_{n,k} =  \frac{\mathcal{K}+ \beta_r}{\frac{L+\bar{N}}{2}+ \alpha_r+1}.
\label{eqt:Mode_posterior_psi_nk}
\end{equation}

\subsection{Background estimation} \label{subsec:Background} 

Our algorithm assumes known signal counts, which can be obtained after removing background counts from observed histograms. In the presence of obscurants, the background can be non-uniform $b_{n,t,k}$, i.e., in addition to pixels and wavelengths it also depends on time bins related to the depth dimension. Assuming a spatially homogeneous distribution of the obscurant, the background level can be assumed smooth. This means that after downsampling, $y_{n,t,k}^{(L)}$ can be represented by the sum of a smooth function $\hat{b}_{n,t,k}$ and 
a sparse signal due to target reflections. Unmixing these two signals is a common signal processing problem and can be solved using several tools, e.g.,  Robust PCA \cite{CandesACM2011}. In this paper, we only require an approximative  estimate of $\hat{b}_{n,t,k}$ and are more interested in efficient solutions.
More precisely, we assume the background has the same temporal shape for all pixels and estimate this shape as follows
\begin{equation}
\bar{b}_{t,k} = \textrm{median} \left(y_{\mathcal{u}_n,t,k}^{(L)} \right)
\label{eqt:b_t}
\end{equation}  
where $\mathcal{u}_n$ represent the indices of the lowest $10\%$ values of $y_{n,t,k}^{(L)}$ to only consider background and reject signal returns.
Akin to \cite{Gyongy_Optica20}, the noise level of each pixel is estimated using the median as follows  
\begin{equation}
\underline{b}_{n,k} = \textrm{median} \left(y_{n,:,k}^{(L)} \right).
\label{eqt:b_n}
\end{equation}   
The smooth  background is then obtained by 
\begin{equation}
\hat{b}_{n,t,k} = \textrm{max} \left(0, \underline{b}_{n,k} + \bar{b}_{t,k} - \bar{\bar{b}}_{k} \right)
\label{eqt:b_tot}
\end{equation}  
with $\bar{\bar{b}}_{k} =   \sum_t {\bar{b}_{t,k}} / T$.
Knowing the background level, the approximate signal counts can be extracted as follows    
\begin{equation} 
{}^s\! y_{n,t,k}^{(\ell)}= \textrm{max}(y_{n,t,k}^{(\ell)}- \hat{b}_{n,t,k} , \, 0)   , \forall n
\end{equation}  for $t \in [t_l ,t_h]$; where $t_l = \textrm{max}(1, {d}^{\textrm{ML}, (\ell)}_{n} - I^l_k)$, $t_h = \textrm{min}(T, {d}^{\textrm{ML}, (\ell)}_{n} + I^r_k)$, where $I^l_k$ and $I^r_k$ represent the attack and trailing width of the $k$th SIR.


\subsection{Stopping criteria} \label{subsec:Convergence_diagnosis}
Two criteria are considered to stop the iterative coordinate decent algorithm for depth and reflectivity. The  first  is  maximum number of iterations. The second evaluates the estimated parameter values and stops the algorithm if the following condition is satisfied
\begin{equation}
 \left\| \bsx^{(i+1)}-\bsx^{(i)}   \right\|_1 \leq  \xi \left( \left\|\bsx^{(i)}\right\|_1  + \xi\right).
\label{eqt:criteria2}
\end{equation}
where $i$ denotes the algorithm iterations and $\xi=0.001$ is a threshold. 

\section{Results on simulated data} \label{sec:Results_on_simulated_data}

This section evaluates the proposed algorithm on simulated data. The section first introduces comparisons algorithms and evaluation criteria. Then we analyse the robustness of the proposed algorithm with respect to sparse and high-background regimes and compare it on a single-wavelength 3D Lidar data. Finally, we generalize the analysis to multiple wavelengths scenarios. All simulations have been performed on a Matlab R2020a  on a computer with Intel(R) Core(TM) i7-4790
CPU@3.60GHz and 32GB RAM. 
\begin{figure*}
\centering
\includegraphics[width=1.95\figwidth]{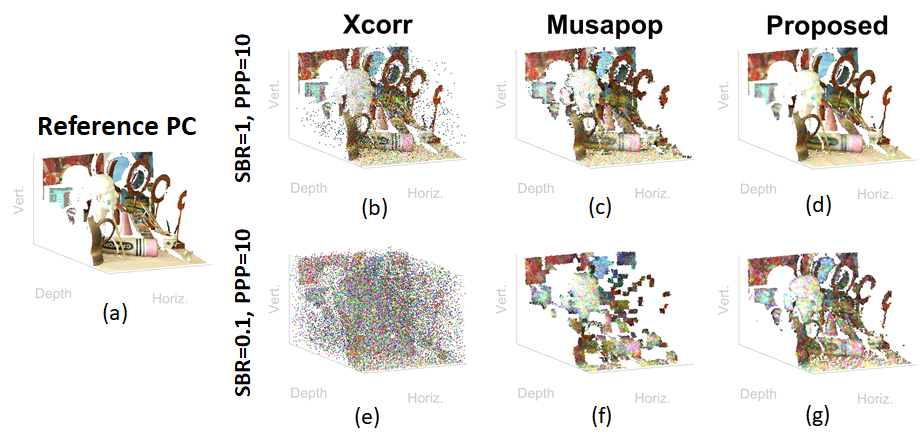}
\caption{3D representations of the art scene for two levels of SBR and PPP obtained with uniform background. (a)
Reference point cloud, (b-e) Xcorr., (c-f) MUSAPOP, (d-g) proposed algorithm. 
 (b-c-d) SBR=1, PPP=10, (e-f-g) SBR=0.1, PPP=10.
 } \label{fig:Art_PC_Ref_3Algos}
\end{figure*} 

\subsection{Comparison algorithms and evaluation criteria} \label{subsec:Algos_Criteria}

To highlight the robustness and benefit of the proposed algorithm, it is  compared to several state-of-the-art algorithms including:
\begin{itemize} 
\item The unmixing algorithm (UA) \cite{Rapp_TCI2017}: considers multi-scale information for robust reconstruction of depth and reflectivity images. It assumes the presence of one surface on all pixels, and is used when analysing robustness to noise and photon-sparse regime imaging. 
\item The RT3D algorithm \cite{Tachella_NC_2019}: assumes the presence of multiple surfaces per-pixel and is used when analysing robustness to noise and photon-sparse regime imaging. 
\item The MUSAPOP algorithm \cite{Tachella_TCI2020}: assumes the presence of multiple surfaces per-pixel and is used when analysing multi-spectral Lidar data.  
\item The Class. algorithm: estimates ${\bsd}^{\textrm{ML}}, {\bsr}_k^{\textrm{ML} }, \forall k$ as in \eqref{eqt:D_ML}, \eqref{eqt:r_ML} from the observed histograms (without removing background)
\item The Xcorr. algorithm: estimates background level  as in \eqref{eqt:b_tot}, then estimates ${\bsd}^{\textrm{ML} }, {\bsr}_k^{\textrm{ML} }, \forall k$ as in \eqref{eqt:D_ML}, \eqref{eqt:r_ML}, respectively (see lines 5-6 in Algo. \ref{alg:Denoising_part}).
\end{itemize}

Comparison results will be analysed qualitatively (by showing reconstruction scenes) and quantitatively using several criteria. The depth performance is measured based on the depth absolute error (DAE) measure DAE$ =  \frac{1}{N'}  \left|\left|\bsd^{\textrm{ref}}- \bsd^{\textrm{est}}  \right|\right|_1$, where $N'$ represents the number of pixels having a target, and $\bsd^{\textrm{ref}}$ and $\bsd^{\textrm{est}} $   are the reference and estimated depth maps with a target, respectively. Similarly,  intensity is evaluated using the intensity normalized absolute error IAE$ =  \frac{\left|\left|\bsr^{\textrm{ref}} - \bsr^{\textrm{est}} \right|\right|_1 }{\left|\left|\bsr^{\textrm{ref}}  \right|\right|_1}$. In addition, we consider the metrics used in \cite{Tachella_TCI2020} to evaluate point clouds.  More precisely,  we consider the percentage of true detections as a function of the distance $\tau$, where a true detection occurs if an estimated point of a given $n$th pixel has a reference point in its surrounding such that $|\hat{d}_n - d_n^{\textrm{ref}} |  \leq  \tau$.  The sum of the estimated points that can not be assigned to any true point at a distance of $\tau$ are considered as false detections. Average normalized IAE is considered for intensity, where pixels with no or false detections are assumed to introduce an error of  $ \frac{r^{\textrm{ref}}_n   }{\left|\left|\bsr^{\textrm{ref}}  \right|\right|_1}$

\subsection{Robustness to sparsity or background counts} \label{subsec:Robustnes_to_sparsity_background_counts}

\begin{figure}
\centering
\includegraphics[width=0.95\figwidth]{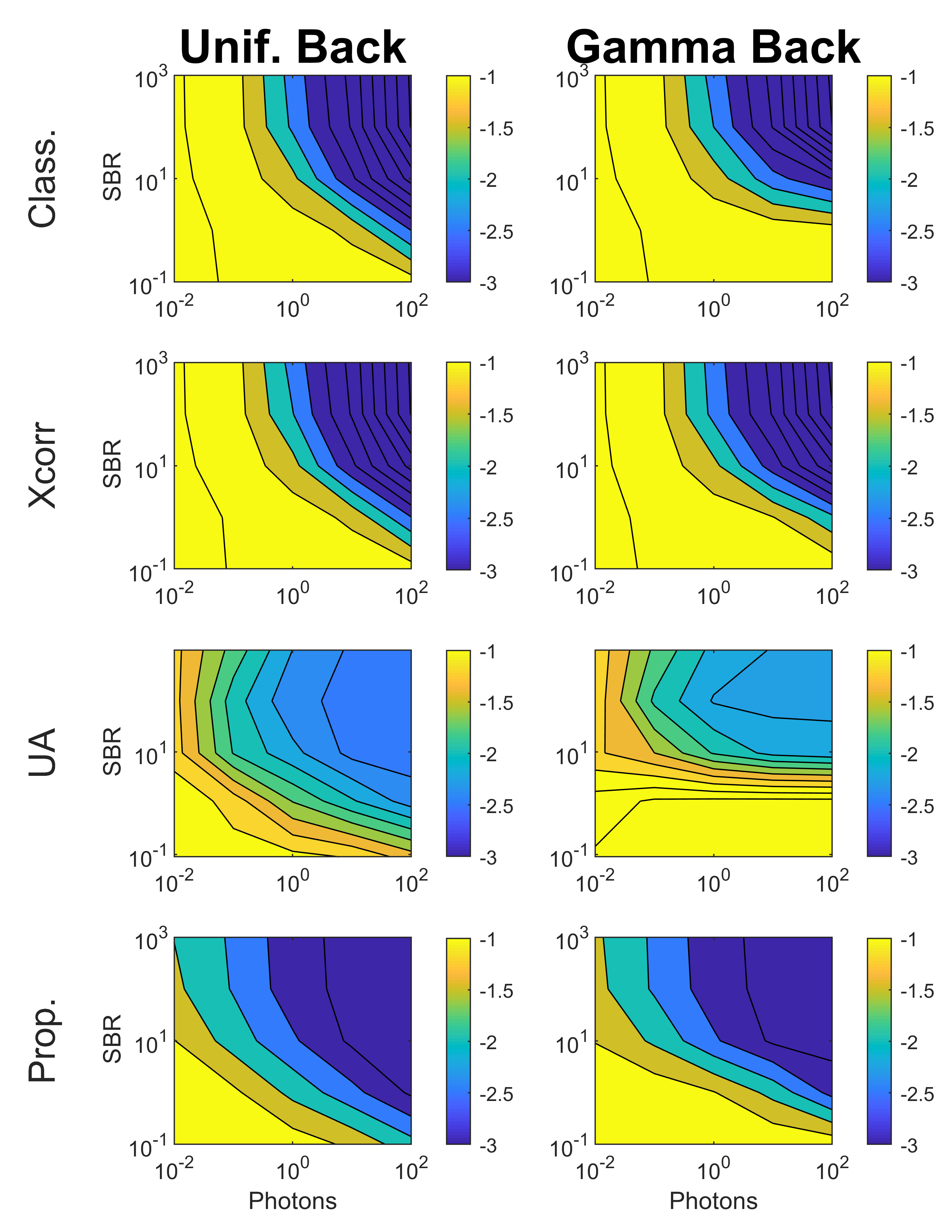}
\caption{Depth absolute errors (in log scale) obtained for the art scene with different algorithms w.r.t. SBR and PPP levels. (top-row) Class. algorithm, (second row) xcorr. algorithm, (third row) UA algorithm, (fourth row) proposed algorithm. (Left-column) data with uniform background, (right-column) data with gamma background. The lower DAE the better. } \label{fig:Art_DAE_Xcorr_UA_Prop_SBR_ppp}
\end{figure}

\begin{figure}
\centering
\includegraphics[width=0.95\figwidth]{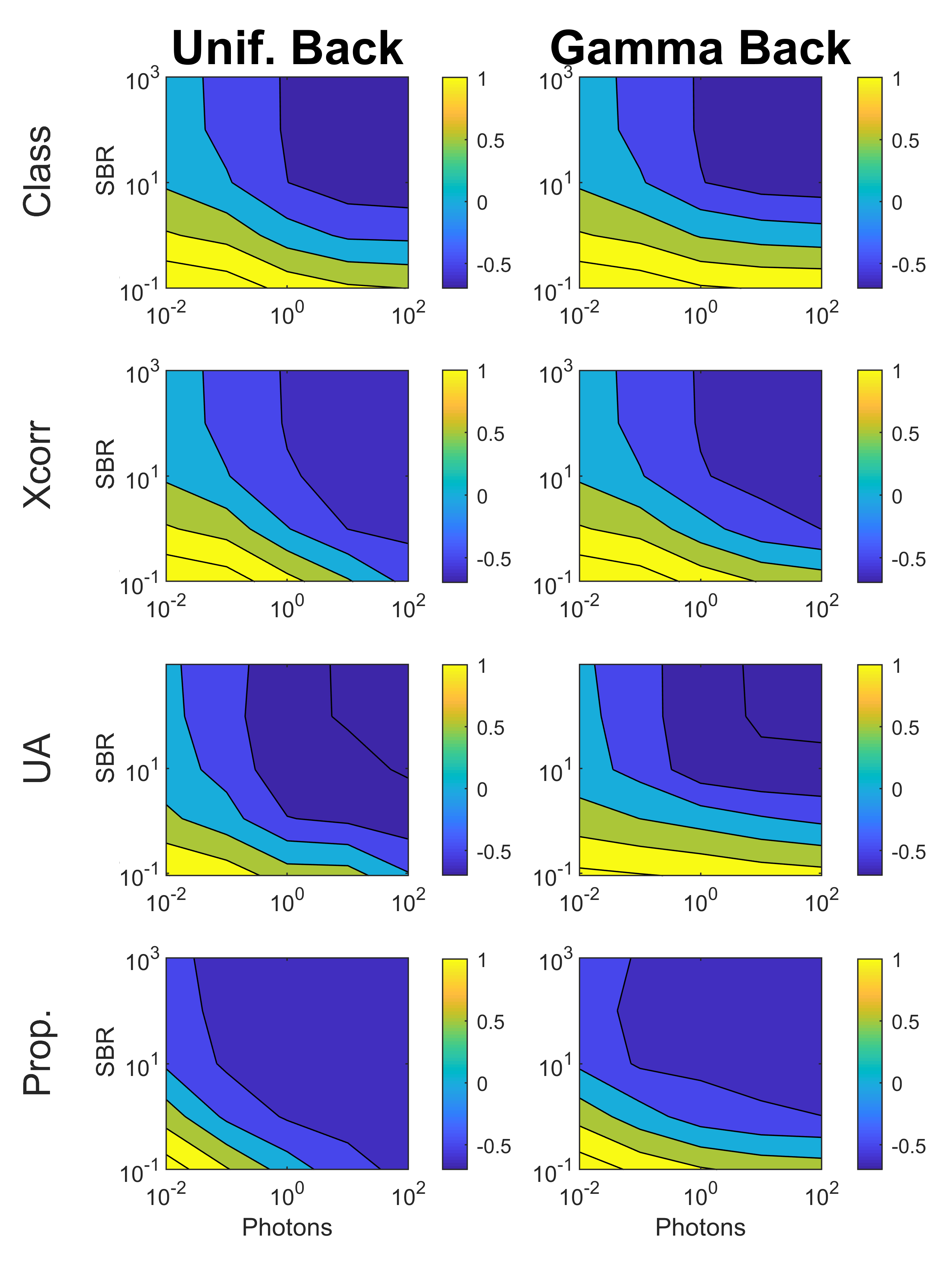}
\caption{Normalized intensity absolute errors  (in log scale) obtained for the art scene with different algorithms  w.r.t. SBR and PPP levels. (top-row) Class. algorithm, (second row) xcorr. algorithm, (third row) UA algorithm, (fourth row) proposed algorithm. (Left-column) data with uniform background, (right-column) data with gamma background. The lower IAE the better.
 } \label{fig:Art_IAE_Xcorr_UA_Prop_SBR_ppp}
\end{figure}

\begin{figure}
\centering
\includegraphics[width=0.95\figwidth]{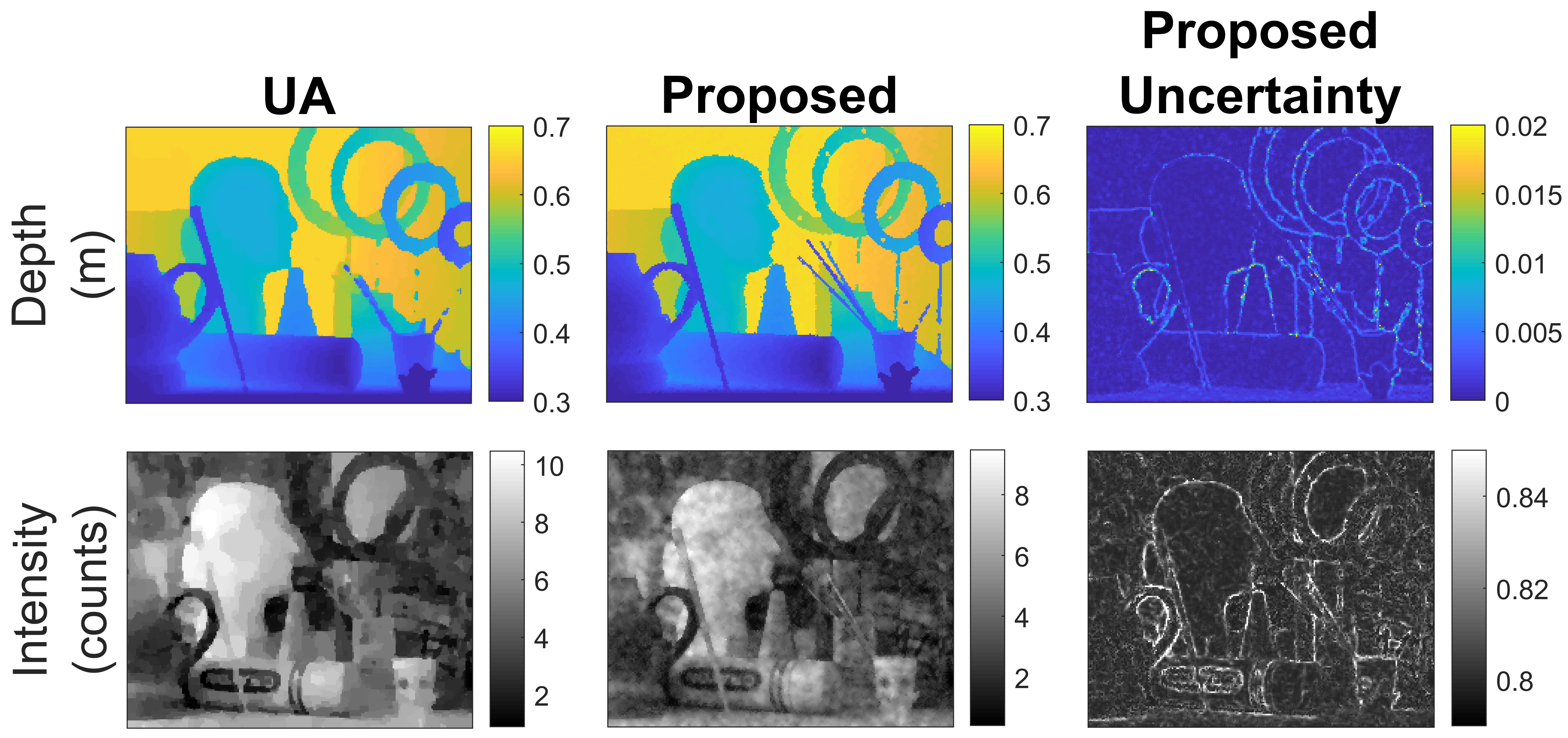}
\caption{Estimated depth and reflectivity maps with the UA and proposed algorithms for SBR=1 and PPP=10 photons  and uniform background. The proposed algorithm provides additional uncertainty maps. } \label{fig:Art_Gray_D_R_UA_Proposed_SBR1_PPP10}
\end{figure}

This section evaluates the algorithm performance under different cases, including the photon sparse regime (low average photon-per-pixels) and low signal-to-background level (SBR), where average SBR$ = \frac{\sum_{n=1}^{N}{r_n}} { \sum_{n=1}^{N}{T b_n} }$. 
Simulations are performed on the realistic Art scene extracted from the Middlebury dataset\footnote{Available in: http://vision.middlebury.edu/stereo/data/}, as it is a cluttered scene used to evaluate many algorithms  \cite{Scharstein_CCVPR_2007,Rapp_TCI2017} (see Fig. \ref{fig:Art_PC_Ref_3Algos} (a)). 
An intensity image is first constructed using the  luminance of the RGB image. The  $283 \times 183$ depth and intensity images are then used to generate a 20ps time bin histograms of counts as in \eqref{eqt:Statistical_model}, while considering a real system impulse response (leading-edge of 3 bins and trailing-edge of 26 bins). The resulting cube of histograms is of size $283 \times 183$ pixels and $T=300$ time bins. To investigate several scenarios, we generate multiple histogram cubes by varying the SBR level logarithmically in $[0.01, 100]$ and the average photons-per-pixel (PPP) in $[0.1, 1000]$ (this PPP combines signal and background counts, useful signal counts can be deduced from the PPP and SBR level). In addition, we consider two background shapes, a conventional uniform shape (i.e., $b_{n,t,k} =b_{n,k}$) where the background level is the same for all time bins, and a gamma shaped background (i.e., $b_{n,t,k} = \mathcal{G} (\alpha,\beta) $ where $\mathcal{G}$ denotes a gamma distribution with parameters $\alpha=2$ and $\beta=30$) often encountered when imaging through obscurants (such as underwater or through fog \cite{satat2018ICCP}).
The proposed algorithm is considered with the following parameters $L=3$ with $q^{(2)} = 3 \times 3$ and $q^{(3)} = 9 \times 9$,  $\zeta=2.7$cm (i.e., 9 time bins),  while considering the first depth and intensity guides (GD1 and GI1).  It is compared with the Classical and Xcorr algorithms (matched filter before and after removing non-uniform background), and the UA algorithm whose depth and intensity regularization parameters were tuned to provide best DAE performance. Fig. \ref{fig:Art_DAE_Xcorr_UA_Prop_SBR_ppp} shows the log scale DAE performance of the considered algorithms when considering uniform (left column) and gamma shaped backgrounds (right column). All algorithms show good results for high SBR and PPP and the performance degrades when decreasing SBR or PPP or when considering a non-uniform background. The proposed algorithm is more robust as it shows the lowest DAE even for extreme cases (DAE$\approx 0.01$ for SBR=1 and PPP=1 photons). The UA algorithm presents second best results, and shows robust results. However, performance is slightly reduced for high SBR and PPP levels due to considering a Gaussian IRF instead of the asymmetric one used to simulate the data. The Xcorr algorithm is more robust than Class. which highlights the importance of the background removal step.
 Fig. \ref{fig:Art_IAE_Xcorr_UA_Prop_SBR_ppp} shows similar behaviours when considering the recovered intensity images, i.e.,  best robustness by the proposed algorithm followed by the UA algorithm. While all algorithms perform well for high SBR and PPP levels, it is worth noting that UA presented best IAEs in this case although its results tend to be over-smoothed (see Fig. \ref{fig:Art_Gray_D_R_UA_Proposed_SBR1_PPP10}). 

In addition to depth and intensity maps, the proposed algorithm also provides their corresponding uncertainty maps (variance of the estimates), which help with decision making. Fig. \ref{fig:Art_Gray_D_R_UA_Proposed_SBR1_PPP10}  shows the depth and reflectivity  maps together with their uncertainty maps  for SBR=1 and PPP=10 photons for uniform background (i.e., 5 signal photons on average). It is observed that the proposed algorithm provides sharp depth maps due to to the use of $\ell_1$ based sparsity inducing prior. 
Note that higher uncertainty is observed on low reflectivity areas, near object edges and on corrupted regions due to high-background levels.   

\begin{figure}
\centering
\includegraphics[width=0.95\figwidth]{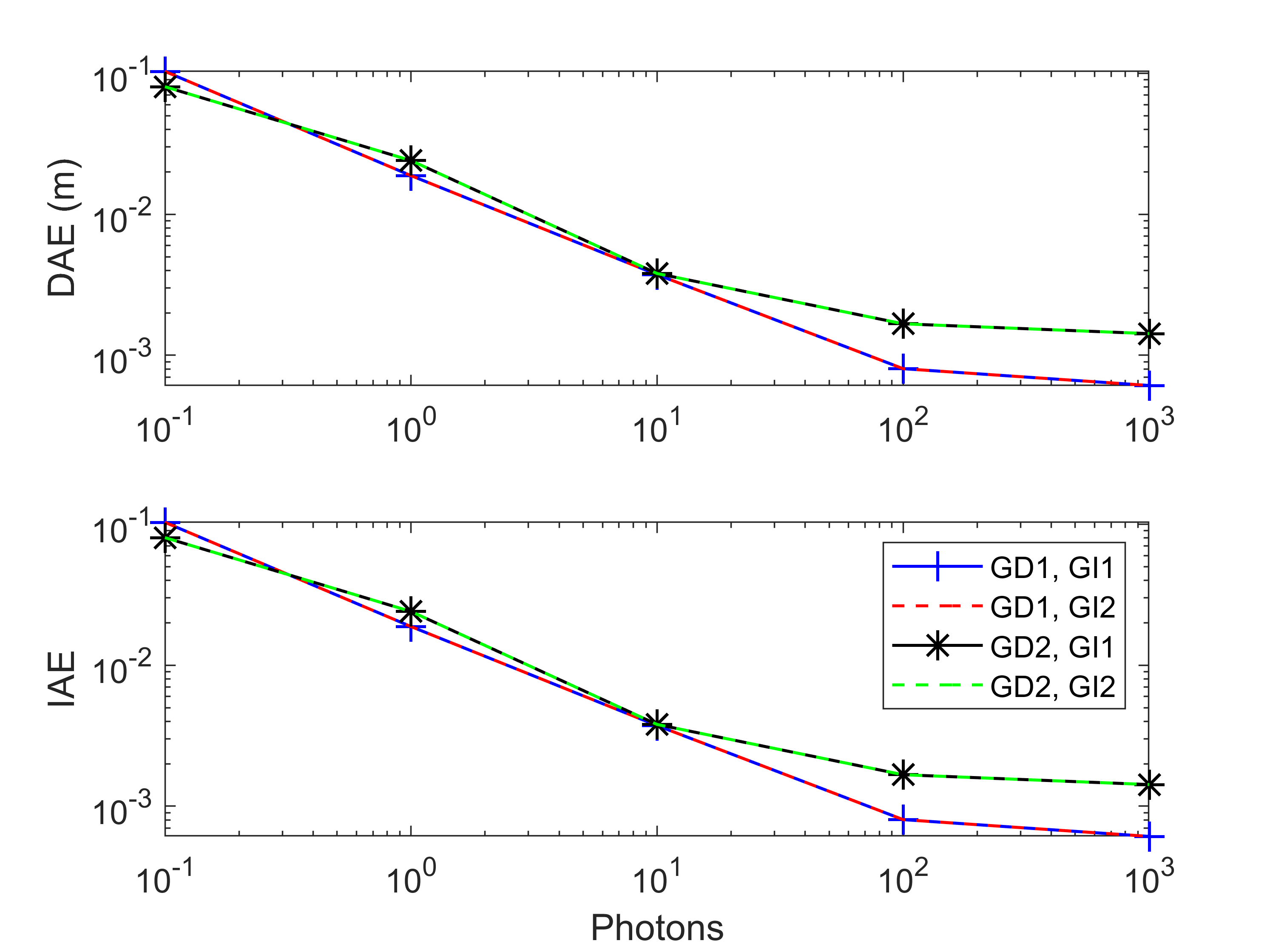}
\caption{Proposed algorithm using different depth and intensity guides on the art scene at SBR=1 and different ppp levels. (top) depth errors, (bottom)  normalized intensity errors. GI3 provides similar results as GI1, GI2 and is not represented for clarity.} \label{fig:Art_DAE_IAE_Prop_DifferentGuides_SBR1}
\end{figure}

An advantage of the proposed algorithm is that it can benefit from state-of-the-art algorithms and use their results as a guide.  We investigate here the performance of the proposed algorithm when considering two depth guides (GD1 and GD2) and three intensity guides (GI1, GI2, GI3).  We repeat the same experiment as above while fixing SBR=1 and varying PPP. Fig. \ref{fig:Art_DAE_IAE_Prop_DifferentGuides_SBR1} shows the DAE and IAE performance of the four variants, indicating an overall similar performance with a slight advantage for GD1 when compared to GD2. GI3 provides similar results as GI1, GI2 and is not represented for clarity. In what follows, we consider the GD1 and GI1 guides for all experiments.

Table \ref{tab:Times_Algos_Art} finally reports the computational time of the algorithms considered indicating fast performance when compared to the UA algorithm. Note that most computations of the proposed algorithm are wasted when building downsampled cubes necessary for background estimation, which are indicated by Xcorr times. 
It should be also noted that most operations of the proposed algorithm are pixel-wise or bin-wise independent, and a much reduced computational time is expected by using parallel computing tools. 
 
\begin{table}  \centering
\centering \caption{Average computational time (in seconds) of the compared algorithms on $283 \times 183 \times 300 \times K$ data cube generated with a uniform and gamma background, where $K$ represents the number of wavelengths. }
\begin{tabular}{|c|c|c|c|c|c|c|}  
   \cline{3-7} \multicolumn{2}{c|}{}  & \multicolumn{5}{|c|}{Average photons per pixel (PPP)} \\
	\cline{3-7} \multicolumn{2}{c|}{}   & $0.1$ & $1$ & $10$ & $100$ & $1000$ \\
\hline  
Art,         &  Class.  &   $  0.4 $ &  $0.4 $ & $0.4 $ & $ 0.4 $ & $ 0.4 $   \\
K=1          &  Xcorr.  &   $ 2.3 $ &  $2.4 $ & $2.4 $ & $ 2.4 $ & $ 2.4 $   \\
wavelength   &  UA      &    $ 37 $ &  $42 $ & $65 $ & $ 62 $ & $ 136 $   \\
             &  Prop.   &    $ 4.2 $ &  $4.2 $ & $3.9$ & $ 3.8 $ & $ 3.8 $   \\  
\hline 
\hline 
Art,        & Class.   &   $ 10 $ &  $10 $ & $10 $ & $ 10 $ & \multicolumn{1}{|c}{}   \\   
K=3,        & Xcorr.   &   $ 17 $ &  $18 $ & $17 $ & $ 17 $ & \multicolumn{1}{|c}{}   \\   
wavelengths         & MUSAPOP   &  $ 548 $ &  $628 $ & $840 $ & $ 1169 $ & \multicolumn{1}{|c}{}    \\   
   & Prop.   &   $ 21 $ &  $21 $ & $19 $ & $ 19 $ & \multicolumn{1}{|c}{}    \\  
\cline{1-6}
 \end{tabular}
\label{tab:Times_Algos_Art}
\end{table}

\subsection{Evaluation on multispectral 3D Lidar data} \label{subsec:Evaluation_ on_multispectral_3D_Lidar_data}
  
\begin{figure}
\centering
\includegraphics[width=1.05\figwidth]{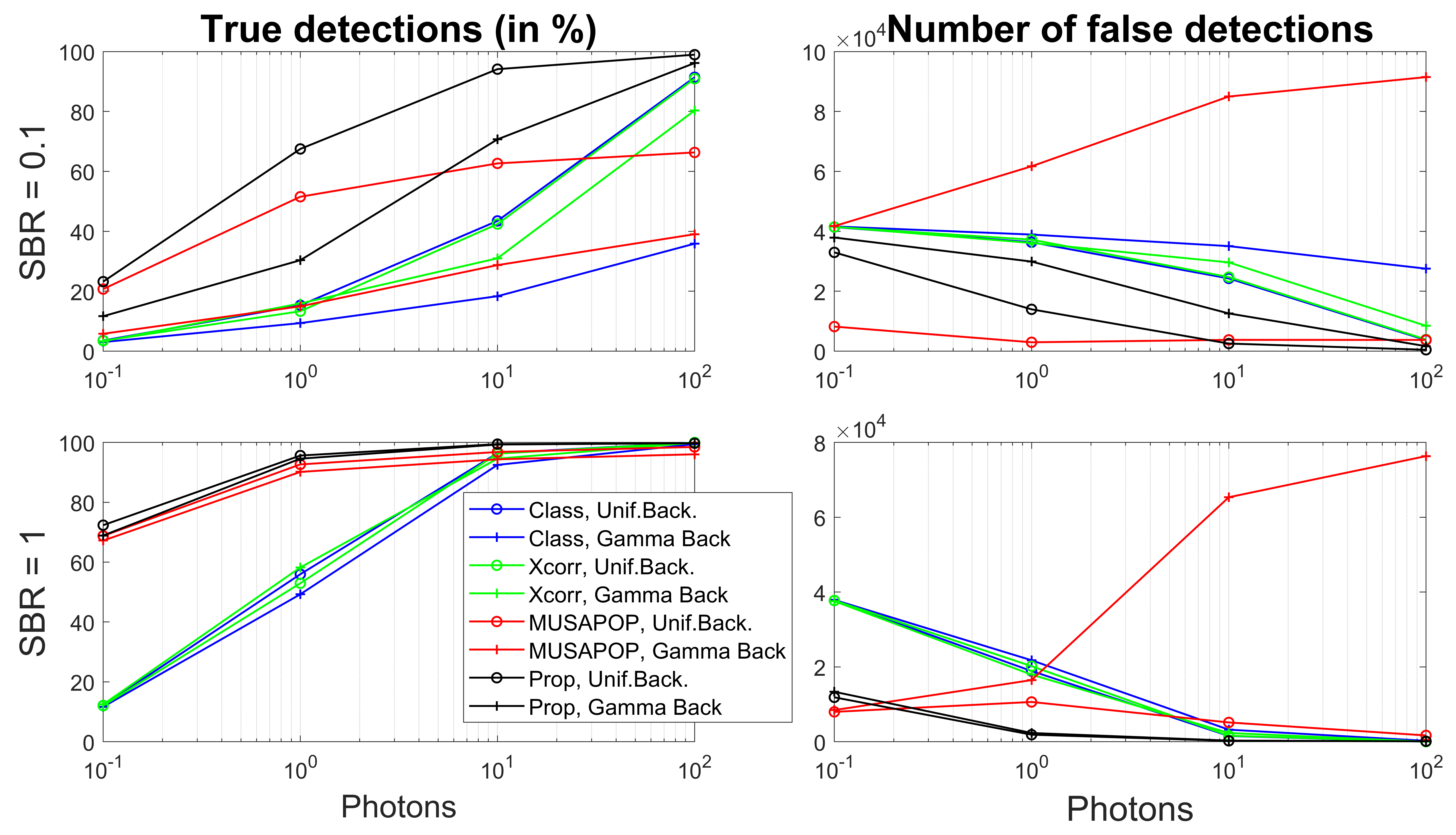}
\caption{PD and false detections of the Class, Xcorr, Musapop and proposed algorithms for different SBR, PPP levels and background shapes, with an error distance of $\tau = 10$bins. } \label{fig:PD_PF_Xcorr_MUSAPOP_Proposed}
\end{figure}

This section analyses the performance of the proposed algorithm  for multi-spectral Lidar imaging. We consider 3 wavelengths of the Art scene to generate three histograms  of counts with the same realistic IRF as in Section \ref{subsec:Robustnes_to_sparsity_background_counts} while varying SBR and PPP levels. The proposed algorithm is considered with the same hyperparameters (as in Section \ref{subsec:Robustnes_to_sparsity_background_counts}), and is compared with the MUSAPOP algorithm (used with the authors' parameters). The latter is designed to process multi-spectral data and delivers point clouds hence the use of probability of detection, and number of false detections to evaluate performance. 
Fig. \ref{fig:PD_PF_Xcorr_MUSAPOP_Proposed} represents these two criteria for four algorithms when considering a uniform  (cross marker) and gamma shaped backgrounds (circle marker) for two SBR levels and several PPPs at $\tau=10$ bins. The proposed algorithm presents best performance highlighting its robustness due to the efficient use of the multi-scale information. Fig. \ref{fig:Art_PC_Ref_3Algos} shows an example of the obtained point clouds with different algorithms for uniform background. MUSAPOP is better than Xcorr, but fails to process the noisy case where only one signal photon is present against 9 background counts on average (SBR=0.1, PPP=10 photons).
The proposed algorithm presents best performance and is robust to noise. In addition, it does not join disconnected surfaces (due to the use of sparsity inducing Laplace prior for the depth) and shows sharp intensity values (due to the weighted and depth guided reflectivity reconstruction). Table \ref{tab:Times_Algos_Art}   finally highlights the fast  computational time of the proposed algorithm when compared to the MUSAPOP algorithm.  

\section{Results on real 3D underwater data} \label{sec:Real_3D_underwater_imaging}

\begin{figure*}
\centering
\includegraphics[width=1.95\figwidth]{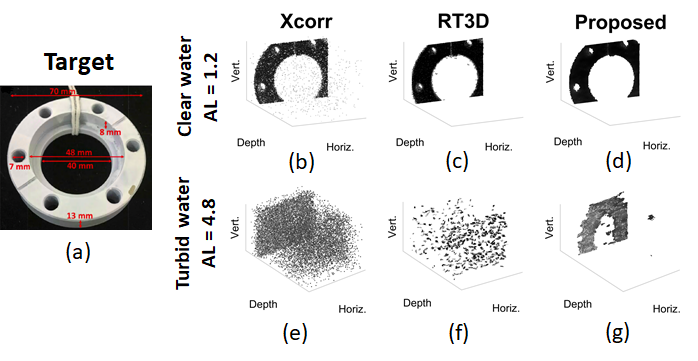}
\caption{(a) Target used for underwater imaging experiments. 3D representations of underwater scenes with (b-e) Xcorr, (c-f) RT3D and (d-g) the proposed algorithm. (b-c-d) clear water with AL=1.2, (e-f-g) turbid water with AL=4.8 (these AL values are for transceiver to target, and not round-trip).  } \label{fig:UW_PC_Xcorr_RT3D_Proposed}
\end{figure*}
\begin{figure}
\centering
\includegraphics[width=0.95\figwidth]{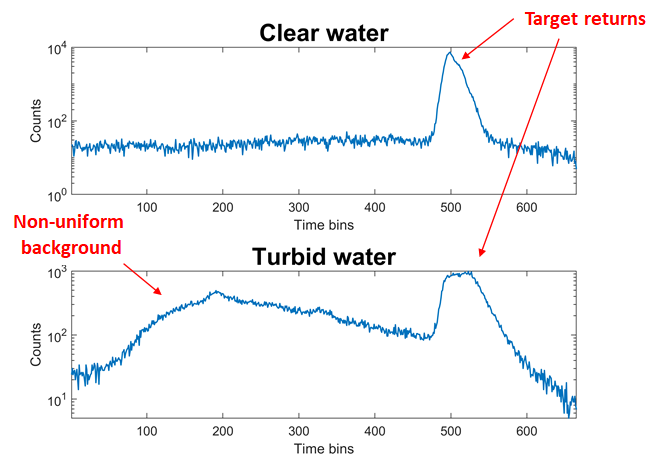}
\caption{Examples of the obtained histogram summed over all pixels for (top) clear water with laser power 1.2 mW, (bottom) turbid water with laser power 8 mW, when imaging the flange target located around the 500 time bin. The bottom curve highlights a non-uniform background (see the region 100 - 400 bins). The shape of the peak is different in the two histograms because the target return is visible in different pixels during the two measurements, as the target moves across the field of view of the camera.   }
 \label{fig:UW_Histograms_CleanTurbidWater}
\end{figure}

The proposed algorithm is validated on real underwater Lidar data of a moving target (painted metal flange
13 mm thick, diameter of 70 mm, with 7 mm diameter holes) put at a stand-off distance of 1.7m from the end of the water tank nearest the sensor (see the target in Fig. \ref{fig:UW_PC_Xcorr_RT3D_Proposed} (a)). 
The data were acquired in  lab settings using a CMOS Si-SPAD detector array based system acquiring binary frames at a rate of $500$fps, with $1$ms acquisition time per frame, $700$ time bins and $34$ps per bin \cite{Maccarone_OE19}.  
The $128 \times 192$ pixels binary frames were pre-processed by building histograms of counts every 10ms (max of 10 counts per histogram). Different concentrations of a commercially available antacid medicine, called Maalox, were mixed with water to obtain varying scattering levels of the imaging environment. With high Maalox concentrations, the turbid water is highly scattering leading to a non-uniform background as shown in Fig. \ref{fig:UW_Histograms_CleanTurbidWater}.

\begin{figure}
\centering
\includegraphics[width=0.95\figwidth]{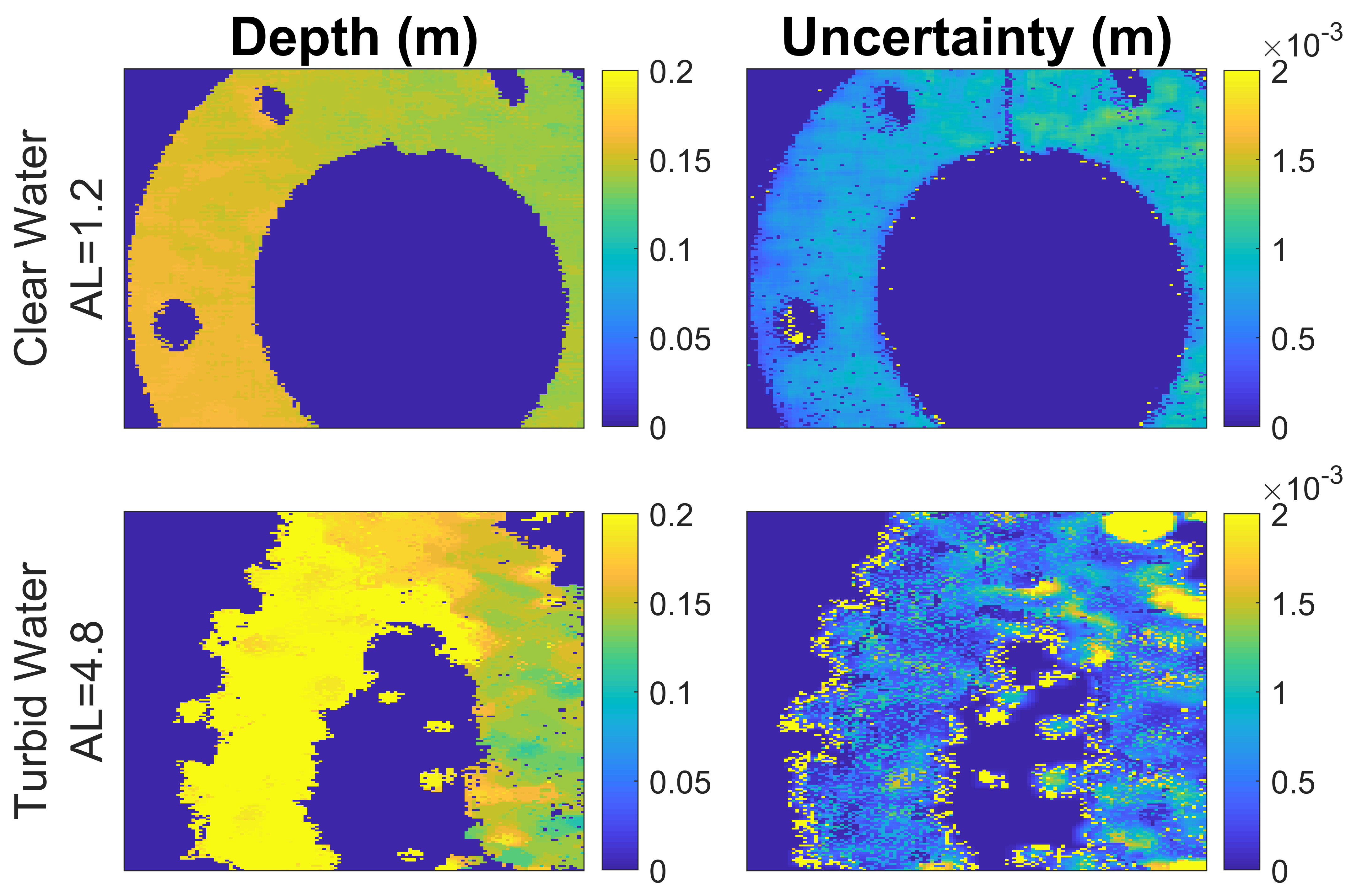}
\caption{Estimated Depth values and uncertainty maps using the proposed method for (top) clear and (bottom) turbid water.   } \label{fig:UW_Uncertainty_Depth_Proposed}
\end{figure}

The proposed algorithm is considered using the hyperparameters of Section \ref{subsec:Robustnes_to_sparsity_background_counts}  and the guides GD1 and GI1.  Results are compared with the Xcorr algorithm and the RT3D algorithm. The UA algorithm is not considered as it assumes the presence of a target in all pixels which is not satisfied in this case. Fig. \ref{fig:UW_PC_Xcorr_RT3D_Proposed} shows the 3D point clouds obtained with the different algorithms for clear water (b-c-d) and turbid water (e-f-g)\footnote{
Attenuation length (AL) is an indication of the effect of optical attenuation, and is the distance over which the  light intensity is reduced to 1/e of its original value. AL = $\alpha d$, when the light propagates a distance $d$ in water with attenuation coefficient $\alpha$ \cite{Maccarone_OE19}.  }. All algorithms performed well in clear water. However, both RT3D and Xcorr performed poorly in turbid water due to non-uniform background affecting the data, and leading to the detection of a fake object in front of the true target. The proposed algorithm successfully eliminates the background counts and retrieve a good reconstruction of the target even under these extreme imaging conditions. In addition, the proposed algorithm also provides uncertainty maps for the estimated parameters, as indicated in Fig. \ref{fig:UW_Uncertainty_Depth_Proposed}. These maps show higher uncertainties when imaging through scattering water,  and near object edges. 
More results when considering other frames at AL=1.2 are provided in   
\href{https://www.dropbox.com/s/vwmufthk9guvuh1/Video_Depth_Reflect_10ms_AL1.2_XcorrW0_Prop.avi?dl=0}{video 1},
\href{https://www.dropbox.com/s/7vmlha5dl5jhkh6/Video_PC_10ms_AL1.2_XcorrW0_Prop.avi?dl=0}{video 2},
and at AL=4.8 in 
\href{https://www.dropbox.com/s/99s63l3dzhx4czo/Video_Depth_Reflect_10ms_AL4.8_XcorrW0_Prop.avi?dl=0}{video 3},
\href{https://www.dropbox.com/s/ievuughgmis1pvg/Video_PC_10ms_AL4.8_XcorrW0_Prop.avi?dl=0}{video 4}.
The supplementary document \cite{HalimiTR2021_RobLidar} shows more results of the proposed algorithm when processing a real multi-spectral Lidar scene. 


\section{Conclusions} \label{sec:Conclusions}
This paper presented a new robust Bayesian algorithm for the reconstruction of multispectral single-photon Lidar data.
The algorithm exploited multi-scale information to improve depth and reflectivity estimates under extreme conditions due to low light level illumination or imaging through turbid media. The approach also provided uncertainty measures regarding the estimates which is crucial for decision making. Results on both simulated and real data highlighted the benefit of the proposed strategy when compared to state-of-the-art algorithms. Future work will  generalize the proposed strategy to process multiple detections per-pixel as observed in object's edges or when imaging through semi-transparent surfaces. Current implementation was done in Matlab, and a computational improvement is expected by using parallel computing tools (such as GPUs) which is being investigated.

\renewcommand{\baselinestretch}{0.92}
\small
\bibliographystyle{IEEEtran}
\bibliography{biblio_all}
\end{document}

%% file: notations.tex
\input com_notations.tex

\newcounter{algo}
\renewcommand{\thealgo}{\arabic{algo}}

%% file: com_notations.tex
\def\bsd{{\boldsymbol{d}}}

\def\bsm{{\boldsymbol{m}}}

\def\bsr{{\boldsymbol{r}}}

\def\bsx{{\boldsymbol{x}}}
\def\bsy{{\boldsymbol{y}}}

\def\bsB{{\boldsymbol{B}}}

\def\bsD{{\boldsymbol{D}}}

\def\bsM{{\boldsymbol{M}}}

\def\bsR{{\boldsymbol{R}}}

\def\bsV{{\boldsymbol{V}}}
\def\bsW{{\boldsymbol{W}}}

\def\bsY{{\boldsymbol{Y}}}

\def\calG{{\mathcal{G}}}

\def\calI{{\mathcal{I}}}